\documentstyle[preprint,aps,epsfig]{revtex}

\begin{document}

\draft
\preprint{
\vbox{
\hbox{ADP-99-39/T376}
\hbox{IU/NTC 99-09}
}}

\title{On the $\xi$-Distribution of Inclusively Produced Particles 
in $e^+e^-$  Annihilation} 

\normalsize
\author{C. Boros, J.T. Londergan$^1$ and  A.W. Thomas}
\address {Department of Physics and Mathematical Physics,
                and Special Research Center for the
                Subatomic Structure of Matter,
                University of Adelaide,
                Adelaide 5005, Australia}

\address{$^1$ Department of Physics and Nuclear
            Theory Center, Indiana University,
            Bloomington, IN 47408, USA}

\date{\today}
\maketitle
\begin{abstract}
We discuss the momentum distributions of inclusively produced 
particles in $e^+e^-$ annihilation.  
We show that the dependence of the position of the maxima of the  
$\xi=\ln(1/z)$ spectra on the mass of the produced particles  
follows naturally  from the  
general definition of fragmentation functions when  
energy-momentum conservation is correctly incorporated. 
\end{abstract}

\newpage 

$e^+e^-$ annihilation provides an excellent opportunity 
to study the fragmentation of quarks into hadrons. 
In particular,  inclusive measurements of particle 
spectra allow us to extract  fragmentation functions 
from such experiments and to test different 
theoretical  models of fragmentation. 
Until now,  fragmentation functions  
have not been  calculated from first principles rather 
they have to be modeled in some way. 
Most of the current approaches   
use different algorithms, such as string and shower algorithms,  
and  model fragmentation of a high energy quark in two phases,  
one of which is purely perturbative, describing the 
radiation and branching of the initial quarks  and 
the other describing the subsequent non-perturbative hadronisation of the 
low energy quarks. 
Here, we follow a different approach. 

Starting from the general definition of the fragmentation functions, 
and explicitly guaranteeing energy-momentum conservation 
we discuss the following interesting property of 
inclusive particle spectra in $e^+e^-$ annihilation. 
When the number of  inclusively
produced particles  is plotted as a function of $\xi=\ln(1/z)$, 
(where $z$ is the momentum fraction  
of the fragmenting quark carried by the produced hadron)
it  exhibits an  approximate Gaussian shape
around a maximum, $\xi^*$.  
The position of the maximum  
depends both on the total centre of mass  energy and on the mass of  
the produced particle \cite{Delphixi,L3xi,Opalxi,SLD}.  
While the shape and the energy dependence of the spectrum can be understood 
in perturbative QCD, as a consequence of the coherence
of gluon radiation \cite{Dok}, the  position of the maximum  is
a free parameter which has to be extracted from  experiment.  
Our main purpose in this paper is to 
show that the dependence of location of the  
maximum on the mass of the produced particle 
follows naturally from the general definition 
of fragmentation functions when  energy-momentum conservation is 
correctly incorporated. 

Our starting point is the general definition of  
fragmentation functions \cite{Collins,Jaffe83,JaffeXi}
\begin{equation}
  \frac{1}{z} D_q (z) 
   =   \frac{1}{4}\sum_n
   \int \frac{d\xi^-}{2\pi}
      e^{-ip^+ \xi^-/z} \, 
         \mbox{Tr} \{ \gamma^+ \, \langle 0|\psi (0) |
 Pn;pp_n  \rangle
        \langle Pn;pp_n)
| \overline{\psi} (\xi^-) | 0\rangle \}.
\label{frag1} 
\end{equation}
(Here, we discuss only the twist two part of the 
unpolarized fragmentation functions.) $\gamma^+$ is defined as  
$\gamma^+ = \frac{1}{\sqrt{2}} (\gamma^0 + \gamma^3)$ and the plus 
components of the momenta are defined as  
$p^+=\frac{1}{\sqrt{2}} (p^0+p^3)$. $p$ and $p_n$ are the 
momenta of the produced particle, $P$, and associated hadronic system $n$.  
Using translational invariance to remove the $\xi^-$ 
dependence in the second the matrix element and the integral
representation of the delta function and projecting out the
plus  components of $\psi$, we obtain
\begin{equation}
  \frac{1}{z} D_q (z)
    =  \frac{1}{2\sqrt{2}} \sum_n
    \delta [(1/z-1)p^+-p_n^+]
        | \langle 0|\psi_{+} (0)
|P n; pp_n   \rangle |^2 .
\label{frag}
\end{equation}
Here, the plus projection is 
defined as $\psi_+ = \frac{1}{2} \gamma_+\gamma_-\psi$. 
Using Eq. (\ref{frag}) 
rather than Eq. (\ref{frag1}) has the
advantage that energy-momentum 
conservation is built in {\it before} any approximation
is made for the states in the matrix element.  
This is similar to  the case of quark distributions   
where the corresponding expression ensures correct support  
of the distribution functions  as 
discussed in Ref. \cite{Signal}.

While the detailed structure of the fragmentation function 
depends on the exact form of the matrix element 
some general  properties follow already from Eq. (\ref{frag}). 
The delta function, for example,  implies that
the  function, $D_q(z)/z$, peaks at
\begin{equation}
   z_{max} \approx \frac{M}{M +M_n}.
\label{max}
\end{equation}
Here, $M$ and $M_n$ are the mass of the produced 
particle and the produced system, $n$, and 
we work in the rest frame of the produced particle. 
Here, we consider the intermediate state as a state 
having a definite mass. (In general, we have to integrate 
over a spectrum of all possible masses.)    
We see that the location of  the maxima of the fragmentation function 
depends on the mass of the system $n$.  While the high $z$ 
region is dominated by the fragmentation of a quark into the 
final particle and a small mass system, large mass systems 
contribute to the fragmentation  at lower $z$ values.

\begin{figure}
\psfig{figure=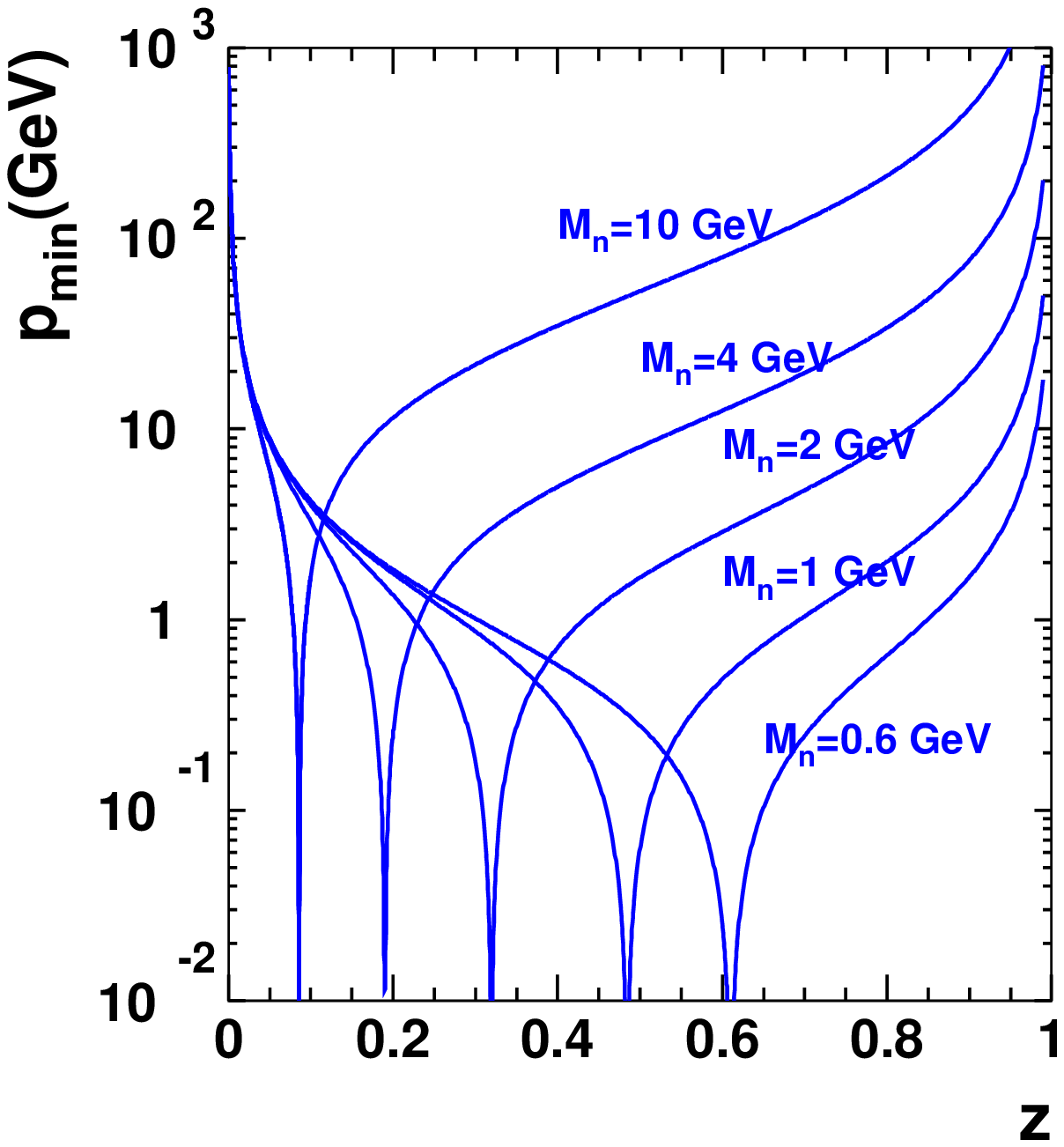,height=14.cm}
\caption{The lower integration limit, $p_{min}$, 
for proton production  as 
a function of $z$ for various masses, $M_n$. } 
\label{fig1}
\end{figure}

We can go a step further and eliminate the $\delta$-function 
in Eq. (\ref{frag})  by integrating over the momentum of the 
unobserved state $n$. 
\begin{equation}
  \frac{1}{z} D_q (z)  =  \frac{1}{2\sqrt{2}} 
 \int^\infty_{p_{min}}
 p_ndp_n |\langle 0|\psi_+ (0) |P n;p p_n 
 \rangle |^2, 
\label{eq:frag}
\end{equation}
with 
\begin{equation}
   p_{min}= |\frac{M^2(1-z)^2-z^2M_n^2}{2M z(1-z)} |.
 \label{max2} 
\end{equation}
The significance of Eqs. (\ref{eq:frag}) and 
(\ref{max2}) is that 
$D_q(z)$ vanishes for both $z\rightarrow 1$ and 
$z\rightarrow 0$. Thus,  fragmentation functions have the 
correct support. 
It is interesting to see how the 
integration region depends on the momentum fraction, 
$z$, for various values of the produced particle system, $M_n$. 
In Fig. \ref{fig1}, we plot 
$p_{min}$ for the production of protons 
as a function of $z$ for different values of $M_n$. 
$p_{min}=0$ gives the value of $z$ at which the contribution of 
a given $M_n$ to the fragmentation function  is largest. 
This gives $z_{max}$ according to Eq. (\ref{max}). 
Further, the region of $z$ where 
the lower integration limit is sufficiently small that the integral 
will be significant,  
becomes narrower with increasing $M_n$. Thus, 
large mass states contribute to the   
fragmentation function at low $z$ and only 
in a very narrow region of $z$.    

\begin{figure}
\psfig{figure=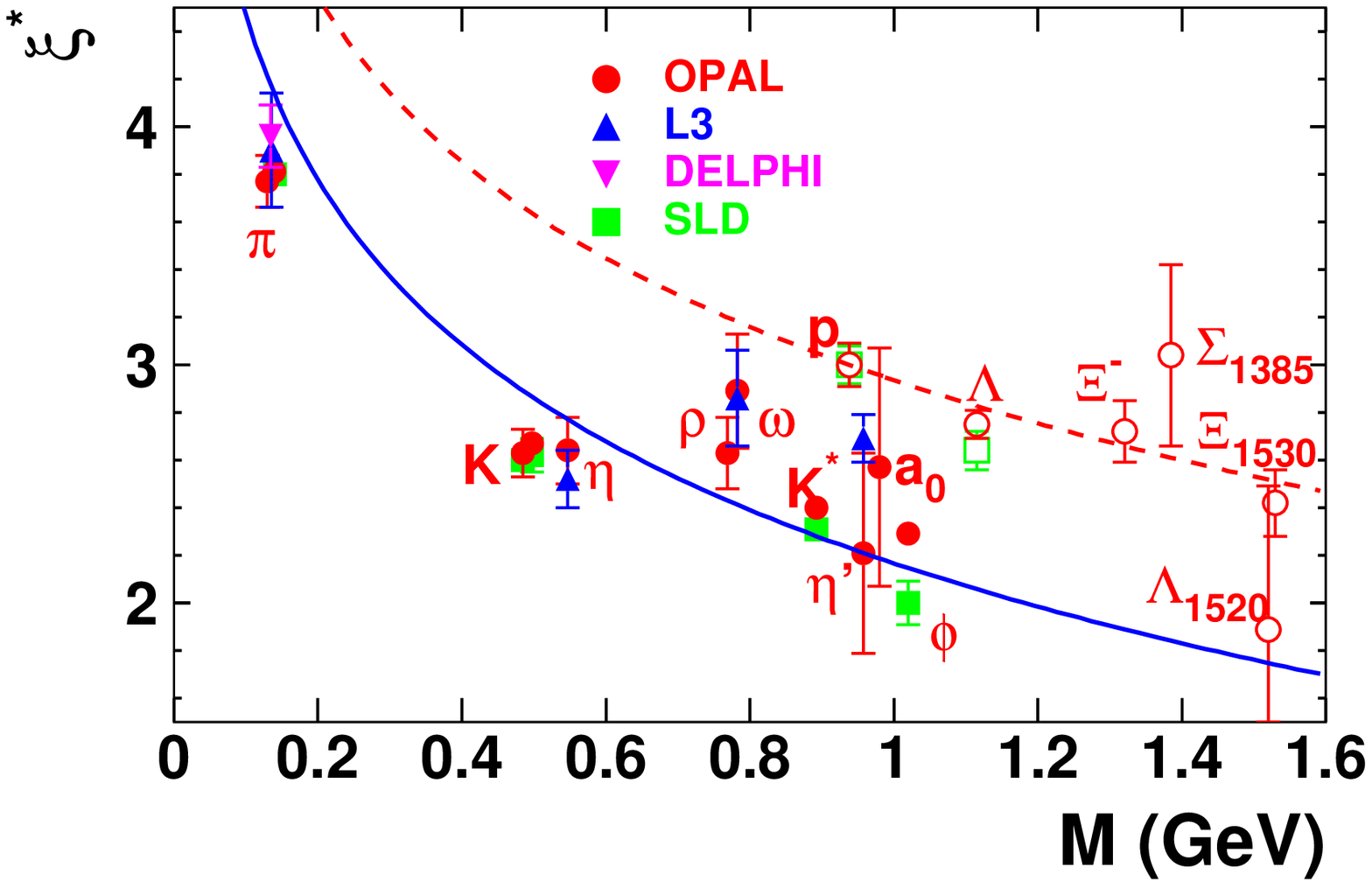,height=10.cm}
\caption{Location of the maxima of the $\xi$ distributions
as a function of the particle mass. The full and open symbols
represent mesons and baryons, respectively. The data are from Refs. 
\protect\cite{Delphixi,L3xi,Opalxi,SLD}. The solid
and dashed lines
are the predictions of Eq. (\protect\ref{maxxi})
for mesons and baryons adjusting the
normalization to $\xi^*_{\eta^\prime}$
and to  $\xi^*_p$, respectively.}
\label{fig2}
\end{figure}

At a given centre of mass  energy, there will
be a maximum value for the mass of the intermediate state which can be
produced in the fragmentation. This maximal mass determines
the ``lower'' edge of the spectrum.   
\footnote{Since the $\xi$-distribution 
is given by  $d\sigma/d\xi= zd\sigma/dz
\sim z D(z)$ it is proportional to $z^2$ times
$D(z)/z$. Although Eq.(\ref{max}) describes the location of
the maximum of the distribution $D(z)/z$
we can expect that Eq.(\ref{max}) is also a good approximation for
the $\xi$-distribution, since the $z$ region where the lower integration 
limit 
($p_{min}$), is sufficient small, is very narrow for large 
masses, $M_n$. Thus, the contributions from a
given $M_n$ to the fragmentation functions are very narrow functions 
in $z$ for large $M_n$. Note, that the 
square of the matrix element in Eq. (\ref{eq:frag}) 
must decrease faster then $1/p_n^2$ in order to give finite result 
for the fragmentation functions. 
Eq. (\ref{max}) gives the maximum of the $\xi$-distribution
exactly in the limiting case when the contribution of a given $M_n$ 
to the fragmentation function is a $\delta$-function.}  
We can use Eq. (\ref{max}) to estimate the maxima
of the $\xi$-distribution associated with this particular 
mass. This maximum determines the
maximum of the fragmentation function in first approximation. 
Although  $M_n$ is not known, it
should be proportional to the available total energy
$E_{CM}$. However, the precise value of $M_n$ is not needed if we are 
only interested  in the {\it relative} position of the maxima of the 
$\xi$ distribution of different particles.  
Taking the difference 
of the maxima of the $\xi=\ln(1/z)$ distributions of  two different particles,  
$a$ and $b$, the dependence on the unknown  value of $M_n$
drops out for sufficiently large $M_n$. It follows from Eq.(\ref{max}) that
\begin{equation}
  \Delta \xi^* = \xi^*_a -\xi^*_b
  \approx \ln \left( \frac{M_a + M_n}{M_b + M_n}\right) +
  \ln \frac{M_b}{M_a} \approx \ln \frac{M_b}{M_a}
\label{maxxi}
\end{equation}
Thus, the difference of the maxima is determined by
the logarithm of the ratio of the masses of the produced particles.  
Since the value of $M_n$ for finite centre of mass  energies 
are in general  different for mesons and baryons,  
Eq. (\ref{maxxi}) will be only valid for the difference of the 
maxima of the mesons or baryons separately.   
We calculated the maxima of the $\xi$ distributions
using this formula and using the maxima of the $\eta^\prime$
and that of the proton distributions as a reference value
for mesons and baryons, respectively. 
The results are compared 
to the experimental data \cite{Delphixi,L3xi,Opalxi,SLD}
in Fig. \ref{fig2}. 
The location of the maxima as a function 
of the mass of the produced particles is reasonably well 
described both for mesons and baryons. 

In conclusion, we have shown that  
the dependence of the position of the maxima of the  
$\xi=\ln(1/z)$ spectra on the mass of the produced particles  
follows from the general definition
of the fragmentation functions and from energy-momentum conservation. 
Our results  are in remarkably good agreement with the data.

\acknowledgments
This work was partly supported
by the Australian Research Council.  One of the authors [JTL] 
was supported in part by National Science Foundation research 
contract PHY-9722706.  One author [JTL] wishes to thank the 
Special Research Centre for the Subatomic Structure of 
Matter for its hospitality during the time this work was 
carried out.

\begin{thebibliography}{9}
\bibitem{Delphixi} W. Adam et al., Delphi Collaboration,
        Z.Phys. C {\bf 69} (1996) 561.
%
\bibitem{L3xi} M. Acciarri et al., L3 Collaboration,
       Phys.Lett. {\bf B328}, 223 (1994);
       {\bf B 371}, 126 (1996);
       {\bf B 393}, 465 (1997) ibid.
%
\bibitem{Opalxi} K. Ackerstaff et al.,
        Opal Collaboration, Eur.Phys.J. {\bf C8}, 241 (1999).
%
\bibitem{SLD} K. Abe et al., SLD Collaboration, Phys.Rev.{\bf D 59},
        052001 (1999).
%
\bibitem{Dok} Yu. L. Dokshitzer, V. A. Khoze, A. H. Mueller and
     S. I. Troyan, Basics of Perturbative QCD, Edition Frontiers, 1991
      and the references therein.
%
\bibitem{Collins}  J. C. Collins and D.E. Soper, Nucl.Phys.{\bf B194}, 445
                     (1982).
%
\bibitem{Jaffe83} R. L. Jaffe, Nucl. Phys. {\bf B229}, 205 (1983).
%
\bibitem{JaffeXi} R. L. Jaffe and X. Ji, Phys.Rev.Lett
                {\bf 71}, 2547 (1993).
%
\bibitem{Signal}  A. I. Signal and A. W. Thomas, Phys. Lett. {\bf B211} and 
  A. W. Schreiber, A. I. Signal, and A. W. Thomas,
 Phys. Rev. {\bf D44}, 2653 (1991).
%
\end{thebibliography}
\end{document}